\documentclass[journal=rsm]{CUP-JNL-DTM}%

\usepackage{graphicx}
\usepackage{multicol,multirow}
\usepackage{amsmath,amssymb,amsfonts}
\usepackage{mathrsfs}
\usepackage{amsthm}
\usepackage{rotating}
\usepackage{appendix}
\usepackage{ifpdf}
\usepackage[T1]{fontenc}
\usepackage{newtxtext}
\usepackage{newtxmath}
\usepackage{textcomp}
\usepackage{xcolor}
\usepackage{lipsum}
\usepackage[colorlinks,allcolors=blue]{hyperref}
\usepackage{booktabs}
\usepackage{array}
\newcolumntype{C}[1]{>{\centering\arraybackslash}p{#1}}
\usepackage{tabularx}
\usepackage{adjustbox}
\usepackage{makecell}
\usepackage{listings}
\lstdefinelanguage{Stata}{
  morekeywords={use, sysuse, clear, gen, egen, replace, drop, keep, if, in, by,
    sort, summarise, sum, tabulate, tab, regress, reg, logit, probit, xtset,
    tsset, foreach, forvalues, while, do, local, global, macro, quietly, noisily},
  sensitive=true,
  morecomment=[l]{*},
  morecomment=[l]{//},
  morecomment=[s]{/*}{*/},
  morestring=[b]"
}

\theoremstyle{definition}

\numberwithin{equation}{section}

\jname{Research Synthesis Methods}
\articletype{RESEARCH ARTICLE}
\jyear{2025}
 
\begin{document}

\providecommand{\expect}{\mathrm{E}}             
\providecommand{\var}{\mathrm{Var}}              
\providecommand{\normaldistn}{\mathrm{Normal}}   
\providecommand{\unifdistn}{\mathrm{Uniform}}    
\providecommand{\halfnormaldistn}{\mathrm{half\text{-}Normal}}   
\providecommand{\lognormaldistn}{\mathrm{log\text{-}Normal}}   
\providecommand{\halftdistn}{\mathrm{half}\text{-}t}   
\newcommand{\logtdistn}{\operatorname{log}\text{-}t}

\providecommand{\rmk}[1]{{[\textit{#1}]}}
\providecommand{\uisd}{\sigma_\mathrm{u}}
\providecommand{\reportingif}{\pi_\star} 
\providecommand{\overallif}{\tilde{\pi}} 

\begin{Frontmatter}

\title[Article Title]{Empirical prior distributions for treatment-by-subgroup interaction heterogeneity in random-effects meta-analysis}

\author[1]{Renato Panaro}
\author[1]{Christian R\"{o}ver}
\author[1,2,3]{Tim Friede}

\authormark{Renato Panaro, Christian R\"{o}ver, Tim Friede}

\address[1]{\orgdiv{Department of Medical Statistics}, \orgname{University Medical Center G\"{o}ttingen}, \orgaddress{\state{G\"{o}ttingen}, \country{Germany}}.\email{renato.panaro@med.uni-goettingen.de}}
\address[2]{\orgname{DZHK (German Center for Cardiovascular Research)}, \orgdiv{partner site Lower Saxony}, \orgaddress{\state{G\"{o}ttingen}, \country{Germany}}}
\address[3]{\orgname{DZKJ (German Center for Child and Adolescent Health)}, \orgaddress{\state{G\"{o}ttingen}, \country{Germany}}}



\abstract{Subgroup analyses are central to the assessment of benefits and risks, where recommendations may depend on evidence that treatment effects differ across patient groups. Valid subgroup claims require evidence based on (within-trial) interaction estimates while accounting for the heterogeneity in those interaction effects.
In the common case of only a few available studies, inference may benefit from the use of prior information on the expected amount of heterogeneity.
Although between-study heterogeneity~($\tau$) has been studied empirically for overall treatment effects, no such calibration exists for treatment-by-subgroup interaction effects. We derive empirical (predictive) prior distributions for overall and interaction effect heterogeneity from over 3{,}000 interaction meta-analyses drawn from the \emph{Cochrane Database of Systematic Reviews (CDSR)}. The resulting effect-measure-specific priors indicate that interaction heterogeneity tends to be substantially smaller than treatment effect heterogeneity. We also show that lower precision of within-trial interaction estimates makes interaction heterogeneity harder to identify. Therefore, the use of empirical priors is particularly valuable in sparse interaction meta-analyses. A motivating example illustrates how priors tailored to interaction effects may substantially improve precision in a meta-analysis compared with standard heterogeneity priors.}
\end{Frontmatter}

\section*{Highlights}
\paragraph{What is already known:}
\begin{itemize}
  \item Investigations of subgroup (interaction-) effects should be based on within-trial interaction estimates
  \item Empirically derived priors for between-study heterogeneity in overall treatment effects are available for several common outcome types and applied for instance in health technology assessments, while such guidance for interactions is lacking
\end{itemize}

\paragraph{What is new:}
\begin{itemize}
  \item We derive empirical prior distributions for both treatment effect heterogeneity  and treatment-by-subgroup interaction heterogeneity using more than 3,000 interaction meta-analyses from the \emph{Cochrane Database of Systematic Reviews (CDSR)}
  \item Interaction heterogeneity tends to be smaller than (overall) treatment effect heterogeneity
\end{itemize}

\paragraph{Potential impact for RSM readers outside the authors’ field:}
\begin{itemize}
  \item Empirical information on heterogeneity of interaction effects (absolute as well as relative to main effects) may have implications for design and analysis in related applications,  such as basket trials, biomarker-stratified and enrichment designs, multi-regional clinical trials, dose-response modelling, and, more generally, hierarchical models
\end{itemize}

\section{Introduction}
\label{sec:Introduction}
Subgroup meta-analyses are central to benefit-risk assessments and health technology assessments (HTA): they assess whether treatment effects differ across patient characteristics such as sex, disease severity, or biomarker status, and inform decisions about whether conclusions on added benefit should differ between patient groups. Current guidance also recommends basing subgroup claims on within-trial contrasts (interaction estimates) rather than on separate analyses based on patient subgroups [\cite{IQWiG2025AllgemeineMethodenV8,FisherEtAl2017}]. Meta-analyses of (main or overall) effect estimates benefit from the inclusion of external evidence on the between-study variability (the \emph{heterogeneity}~$\tau$), in particular when only a few studies are considered [\cite{FriedeRoeverWandelNeuenschwander2017a}].
Assumptions on likely magnitudes of heterogeneity may be motivated by general considerations, or may be backed by empirical data [\cite{RoeverEtAl2021}].

The use of larger collections of previous meta-analyses to derive predictive distributions that may serve as heterogeneity priors has been pioneered by Rhodes \emph{et~al.}~[\cite{RhodesEtAl2015}] and Turner \emph{et~al.}~[\cite{TurnerEtAl2015}], who used the \emph{Cochrane Database of Systematic Reviews (CDSR)}~to motivate prior distributions for binary (logarithmic odds ratio, log-OR) and continuous (standardised mean difference, SMD) outcomes.
R\"{o}ver \emph{et~al.} [\cite{RoeverEtAl2023}] discussed the problem in general terms and formalised the approach that was later implemented by the \emph{Institute for Quality and Efficiency in Health Care (IQWiG)} based on a more specific set of historical meta-analyses [\cite{LilienthalEtAl2024}].
Meanwhile, no empirical work currently addresses heterogeneity of treatment-by-subgroup interaction effects, even though this quantity is important in many regulatory assessments and other applications in personalised medicine [\cite{IQWiG2025AllgemeineMethodenV8}].

A difficulty is that treatment-by-subgroup interactions are typically estimated with substantially lower precision (larger standard errors) than overall treatment effects. This feature is even amplified if subgroups vary in size [\cite{Brookes2004, Riley2022SIM}]. 
Consequently, between-trial heterogeneity on the interaction scale is generally harder to estimate empirically than heterogeneity on the treatment effect scale, while precision gains from the inclusion of external evidence may be expected to be more substantial.

The remainder of the paper is organised as follows.
Section~\ref{sec:Methods} introduces the  methods.
In Section~\ref{sec:Empiricaldata}, empirical prior distributions for heterogeneity in both
overall and interaction effects are
estimated from the CDSR\@. 
Section~\ref{sec:MotivatingExample} presents an application to a motivating example.
Section~\ref{sec:Discussion} concludes with limitations and directions for
future research.

\section{Methods}
\label{sec:Methods}

\subsection{Random-effects meta-analysis of treatment effects}
\label{sec:nnhm}

The \emph{normal-normal hierarchical model (NNHM)} is commonly utilised for meta-analyses of several studies allowing treatment effects to vary across studies.
Each of $k$~studies contributes an effect size
estimate~$y_j$ with a (known) standard error~$\sigma_j$ ($j=1,\ldots,k$). The notation in this work is generic: $y_j$~refers to a treatment effect on a continuous scale, which may also be used as an approximation for discrete data (such as log-ORs based on binary data). $y_j$~is assumed to (approximately) follow a normal distribution centred on a study-specific true effect~$\theta_j$, which is itself drawn from a
population of exchangeable effects,
\begin{equation}
 y_j = \theta_j + \varepsilon_j, \qquad
 \varepsilon_j \vert \sigma_j \sim \normaldistn\!\left(0,\,\sigma_j^{2}\right),
 \qquad
 \theta_j \vert \mu, \tau \sim \normaldistn\!\left(\mu,\,\tau^{2}\right).
 \label{eqn:nnhm}
\end{equation}
The between-trial heterogeneity~$\tau$ is difficult to estimate, especially when data are sparse, and it has a direct impact on the precision of the resulting effect estimate. At one extreme, homogeneity~($\tau = 0$) implies \emph{complete pooling}, i.e., all studies estimate a common effect ($\mu = \theta_1 = \dots = \theta_k$). At the other extreme, no pooling~($\tau \to \infty$) treats study-specific effects as unrelated. Between these extremes, \emph{partial pooling} arises when $\tau$ is small to moderate, allowing trial-specific effects to vary around the overall mean~$\mu$ while still borrowing strength across studies~[\cite{Thompson1994, Viechtbauer2005, Higgins2008, StijnenEtAl2010}].

\subsection{The summarising prior approach}
\label{sec:summarising-prior}

To derive a prior for~\(\tau\) empirically, we can learn it from the heterogeneity observed across a large collection of existing meta-analyses on a comparable outcome scale. Rather than treating each meta-analysis in isolation, we can pool heterogeneity information across past syntheses to inform a hyper-prior on the magnitude of~\(\tau\). To this end, the NNHM from Section~\ref{sec:nnhm} is extended by an additional layer describing a ``population'' of heterogeneity values for a larger number of meta-analyses.

Within the random-effects hierarchical model for a meta-analysis indexed by~\(m\), let \(y_{mj}\) denote the observed effect estimate from trial~\(j\), with reported standard error \(\sigma_{mj}\). We assume that trial-specific underlying effects $\theta_{mj}$ are exchangeable within every meta-analysis $m$, and the hierarchical model is specified as
\begin{eqnarray}
y_{mj} = \theta_{mj} + \varepsilon_{mj}, \quad \varepsilon_{mj} \vert \sigma_{mj} &\sim& \normaldistn\!\left(0, \sigma_{mj}^{2}\right), \quad
\theta_{mj} \vert \mu_m, \tau_m \sim \normaldistn\!\left(\mu_m, \tau_m^{2}\right), \\ \tau_m \vert s &\;\sim\;& \halfnormaldistn(s), \qquad s \;\sim\; \unifdistn(0,\, 10), \label{eqn:summarising}
\end{eqnarray}
where $\mu_m$ denotes the meta-analysis-specific mean effect and $\tau_m$ its between-trial heterogeneity [\cite{Roever2020,RoeverEtAl2023}]. The magnitude of the prior’s upper bound for the scale parameter~$s$ depends on the effect measure under consideration; for instance, a value of~10 is considered suitable for the effect measure scales investigated in the present work. The heterogeneity standard deviation~$\tau$ is assigned a common scale-family density, such as the half-normal distribution in~\eqref{eqn:summarising}, while alternative distributional forms, such as the exponential or log-normal, may also be useful [\cite{RoeverEtAl2023}].

The \emph{summarising (hyper-) prior} approach leverages information from a collection of existing meta-analyses (\(m = 1, \dots, M\)) to derive a \emph{predictive distribution} for a new (``future'') heterogeneity parameter, which may often be simplified to estimating the heterogeneity prior's parameter(s) (here: the half-normal scale parameter~$s$).
Technically, the predictive distribution for between-trial heterogeneity in the new meta-analysis is obtained by integrating the conditional distribution~$p(\tau^\star \vert s)$ (\ref{eqn:summarising})
over the scale parameter's posterior
\begin{eqnarray}
p\!\left(\tau^{\star} \vert y_1, \dots, y_M, \sigma_1, \dots, \sigma_M\right)
 &=& \int p\!\left(\tau^{\star} \vert s\right) \, p\!\left(s \vert  y_1, \dots, y_M, \sigma_1, \dots, \sigma_M \right) \, \mathrm{d}s.
\label{eqn:tau_predictive}
\end{eqnarray}
This (posterior predictive) distribution then serves as a prior for the heterogeneity parameter in a new meta-analysis.
From~(\ref{eqn:tau_predictive}) one can see that the predictive distribution results as a \emph{scale mixture} distribution.
If a half-Normal distribution was used to model heterogeneity parameters (as in~(\ref{eqn:summarising})) and the scale~$s$ is estimated with sufficient precision, a half-Normal distribution may be used to approximate the predictive distribution. On the other hand, if uncertainty in~$s$ is substantial, the half-normal scale mixture might for example be approximated via a heavier-tailed half-Student\mbox{-}$t$ distribution [\cite{RoeverEtAl2023}].

\subsection{Random-effects meta-analysis of treatment-by-subgroup interaction effects}
\label{sec:interaction-nnhm}

We consider a meta-analysis of $k$~studies, each reporting treatment effects for two mutually exclusive and complementary patient subgroups. In study~$j$, let $y^A_{j}$ and $y^B_{j}$ denote the estimated treatment effects for subgroups~$A$ and~$B$, both on a continuous scale, with corresponding standard errors~$\sigma_{Aj}$ and~$\sigma_{Bj}$, treated as known for the purpose of synthesis. The within-trial interaction effect is defined as
\begin{equation}
 g_j := y^B_{j} - y^A_{j},
 \label{eqn:gj}
\end{equation}
which is commonly of interest for identifying treatment effect modification [\cite{IQWiG2025AllgemeineMethodenV8, Berlin2002, FisherEtAl2017, Godolphin2022}].
Under the usual normal approximation, $g_j$~is approximately normally distributed around the true study-specific interaction~$\gamma_j$,
\begin{equation}
 g_j = \gamma_j + \epsilon_j, \qquad
 \epsilon_j \vert \sigma_{g,j} \sim \normaldistn\!\left(0,\,\sigma_{g,j}^{2}\right),
 \qquad
 \gamma_j \vert \gamma, \tau_\gamma \sim \normaldistn\!\left(\gamma,\,\tau_\gamma^{2}\right).
 \label{eqn:interaction-nnhm}
\end{equation}
where  (due to the disjoint subgroups) \( \sigma_{g,j}^{2} = \sigma_{Bj}^{2} + \sigma_{Aj}^{2}\). 
The interaction estimates~$g_j$ are synthesised using the NNHM of Section~\ref{sec:nnhm}, applied to interactions rather than treatment effects.

For ratio measures reported on the log scale, such as the log-OR, log-risk-ratio (log-RR), or log-hazard-ratio (log-HR), the within-trial interaction $g_j = y^B_{j} - y^A_{j}$ is a difference of log ratios, which on the original scale corresponds to a ratio of ratios: the ratio of odds ratios (ROR), ratio of risk ratios (RRR), or ratio of hazard ratios (RHR), respectively. For absolute measures such as the risk difference (RD) or standardised mean difference (SMD), the interaction $g_j$ is a difference of risk differences (DRD) or a difference of standardised mean differences (DSMD).
In all cases, interactions~$g_j$ are estimated \emph{within} each trial by contrasting the subgroup-specific effect estimates.

\subsection{Heterogeneity of interaction effects}
\label{sec:Heterogeneityofinteractioneffects}

Within-trial interaction estimates usually have larger standard errors than overall treatment effects, making both interactions and their heterogeneity harder to estimate. We illustrate this with a simple normal model; similar conclusions hold on other scales, though the exact inflation may depend on event risk~[\cite{Kuss2015}] and allocation ratio [\cite{DumvilleEtAl2005}]. Assuming standard errors are approximately inversely proportional to the square root of sample size, consider study~\(j\) with total size \(n_j\), subgroup-\(B\) proportion \(p_j\), and interaction estimate \(g_j\) as in~\eqref{eqn:gj}. Writing \(n_{Aj}=(1-p_j)n_j\) and \(n_{Bj}=p_j n_j\), the subgroup-specific standard errors are
\begin{equation}
\sigma_{Aj} \approx \frac{\uisd}{\sqrt{(1-p_j)n_j}}
\qquad \mbox{and} \qquad
\sigma_{Bj} \approx \frac{\uisd}{\sqrt{p_j n_j}},
\end{equation}
where $\uisd$ is the \emph{unit information standard deviation (UISD)}, assumed equal in both subgroups [\cite{RoeverEtAl2021}].
Because \(g_j=y_{Bj}-y_{Aj}\) contrasts two independent subgroup estimates, its standard error is
\begin{equation}
\sigma_{g,j}
\;=\; \sqrt{\sigma_{Bj}^{2} + \sigma_{Aj}^{2}}
\;\approx\; \frac{\uisd}{\sqrt{n_j}}\cdot\frac{1}{\sqrt{p_j(1-p_j)}}.
\label{eqn:quadrupleFormula}
\end{equation}
By contrast, the overall treatment effect from all \(n_j\) participants has standard error~\(\frac{\uisd}{\sqrt{n_j}}\). Thus, the treatment effect uses the full sample, whereas the interaction combines two subgroup effects, each based on only part of the data [\cite{Brookes2004}]. 
Precision is maximal when subgroups are balanced ($p_j \approx 0.5$); if one subgroup is smaller, the interaction inherits the smaller subgroup’s imprecision. 

When the~$\sigma_{g,j}$ are large, most variation in~$g_j$ reflects within-trial noise rather than true between-trial heterogeneity, limiting information to estimate~$\tau_\gamma$ [\cite{CochraneHandbookV6}]. Standard frequentist heterogeneity estimators then tend to pile up at zero, giving downward-biased or boundary estimates [\cite{Viechtbauer2005,VeronikiEtAl2015}]. In a Bayesian framework, informative heterogeneity priors can stabilize estimation, avoid overestimating~$\tau_\gamma$, and improve the precision of interaction estimates.

Between-trial heterogeneity in interaction effects depends on the context as well as the scale on which the interaction is defined. As for overall treatment effects, the same amount of between-study variation~($\tau$) may have very different practical implications depending on whether effects are expressed on an absolute scale, such as the RD, or on a relative scale, such as the RR or OR. On log-ratio scales, heterogeneity acts multiplicatively, whereas on absolute scales it acts additively, so numerical values of~$\tau$ are not directly comparable across scales. The same heterogeneity parameter may imply very different ranges of risk probabilities depending on the effect scale, and heterogeneity on absolute and relative scales therefore need to be interpreted differently~[\cite{DeeksAltman2001, RoeverEtAl2021}].

\section{Empirical priors on interaction heterogeneity}
\label{sec:Empiricaldata}
\subsection{Data extraction and organisation}

Using a large set of CDSR subgroup meta-analyses, we estimate effect-measure-specific empirical predictive distributions for treatment-by-subgroup interaction heterogeneity and examine their approximation by simple moment-matched half-normal priors. Starting from the full CDSR, we restricted attention to records reporting interaction meta-analyses and then applied eligibility criteria to retain only endpoint-level data suitable for interaction modelling. In particular, the same set of trials was required to be represented in both subgroup strata and each subgroup in a meta-analysis needed to include at least two eligible studies. Meta-analyses involving more than two subgroups were also excluded. All reviews were downloaded and parsed using the \texttt{cochrane} R~package [\cite{cochrane-package}]. After all filtering steps, the final dataset comprised \(10{,}406\) endpoint pairs from \(3{,}062\) meta-analyses, each reporting exactly two pre-specified subgroups.

Table~\ref{tab:dataset-summary} summarises the empirical results by outcome type and effect measure. It shows that the numbers of meta-analyses and endpoint pairs differ markedly across effect measures, with the largest counts for RR, OR, and SMD and the smallest counts for RD and IRR\@. At the same time, interaction meta-analyses are typically small, with median numbers of studies per meta-analysis ranging from two to three and narrow interquartile ranges. 

\begin{table}[!htbp]
  \centering
  \caption{Descriptive summary of the empirical calibration dataset used to derive outcome-specific empirical prior distributions for heterogeneity. Results are stratified by outcome type and effect measure. For each effect measure, the table reports the number of eligible meta-analyses ($M$), the total number of contributing studies, the total number of patients, and the number of studies included in each meta-analysis ($k$), summarised by the median and interquartile range. Binary outcomes are represented by odds ratios (ORs), risk ratios (RRs), and risk differences (RDs), while time-to-event, rate, and continuous outcomes are represented by hazard ratios (HRs), incidence rate ratios (IRRs), and standardised mean differences (SMDs), respectively}
  \label{tab:dataset-summary}

  \setlength{\tabcolsep}{4pt}
  \renewcommand{\arraystretch}{1.1}

  \resizebox{\textwidth}{!}{%
  \begin{tabular}{@{}lrrrcc@{}}
    \toprule
    outcome scale & & & & \multicolumn{2}{c}{Number of studies per meta-analysis ($k$)} \\
    \cmidrule(lr){5-6}
    & \shortstack{number of\\meta-analyses ($M$)}
    & \shortstack{number of\\studies}
    & \shortstack{number of\\patients}
    & median
    & \shortstack{interquartile\\range} \\
    \midrule
    \multicolumn{6}{l}{\textit{binary outcomes}} \\
    odds ratio (OR) & 346 & 1{,}058 & 1{,}460{,}229 & 2 & 2--4 \\
    risk ratio (RR) & 2{,}025 & 7{,}104 & 18{,}046{,}099 & 2 & 2--4 \\
    risk difference (RD) & 44 & 178 & 71{,}174 & 3 & 2--4 \\[1ex]
    \multicolumn{6}{l}{\textit{other outcomes}} \\
    hazard ratio (HR) & 144 & 429 & 182{,}186 & 2 & 2--3 \\
    incidence rate ratio (IRR) & 28 & 76 & 66{,}081 & 2 & 2--3 \\
    standardised mean difference (SMD) & 475 & 1{,}561 & 362{,}918 & 2 & 2--4 \\
    \bottomrule
  \end{tabular}%
  }
\end{table}
An important distinction from the related work of Turner and Rhodes [\cite{RhodesEtAl2015,TurnerEtAl2015}] is that they considered all eligible meta-analyses, and transformed effect estimates, wherever possible, to common measures (ORs or SMDs). In contrast, we retained the effect measures as originally reported and based the analysis on these.

\subsection{Results}

\begin{figure}[!htbp]
\centering
\includegraphics[width=\textwidth]{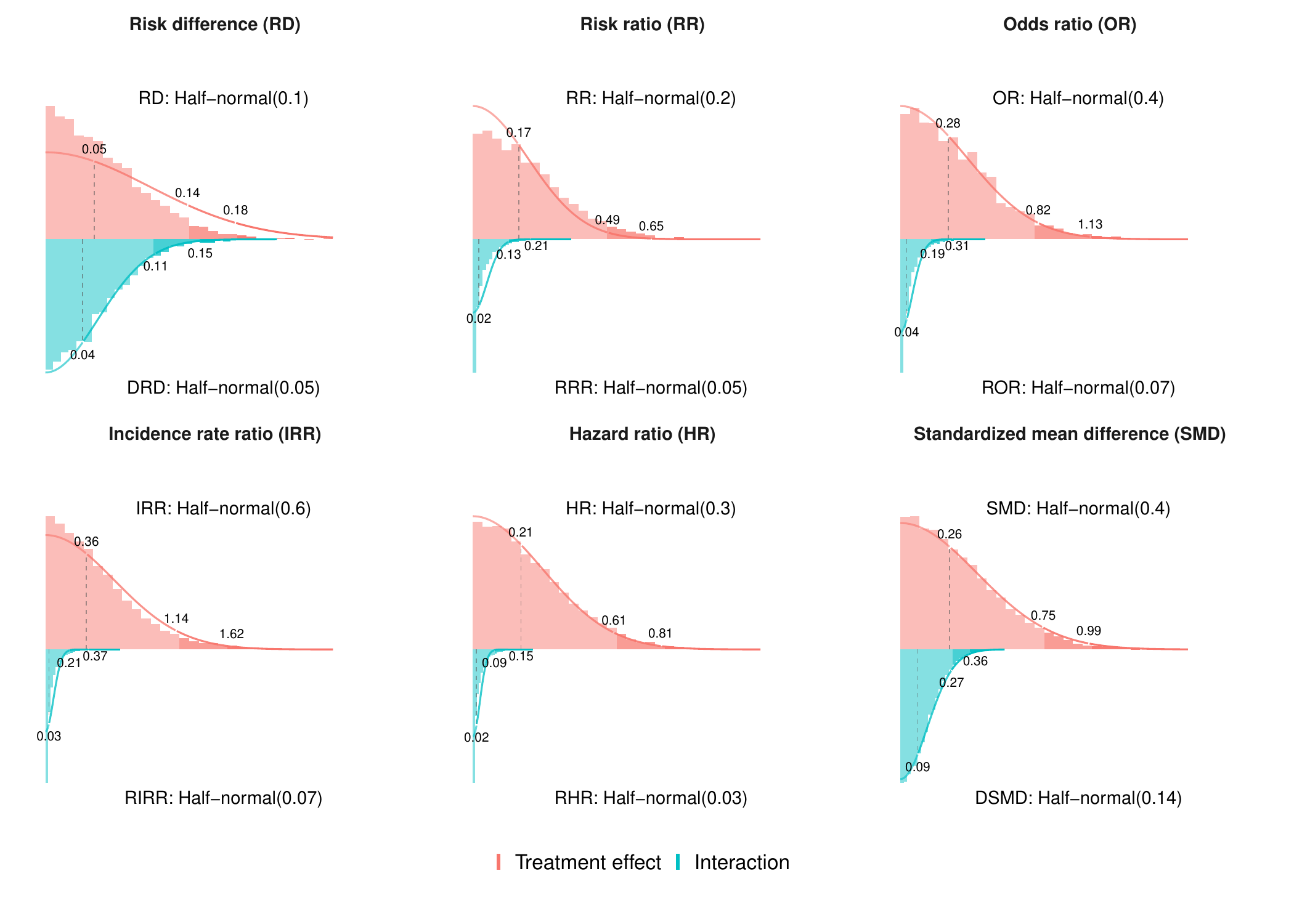}
\caption{Empirical predictive distributions of heterogeneity by outcome type.
For each effect measure, the top panel shows the predictive distribution of
treatment effect heterogeneity~($\tau^\ast$) and the lower panel that of
treatment-by-subgroup interaction heterogeneity~($\tau_\gamma^\ast$).
Posterior predictive histograms are overlaid with half-normal approximations via
moment matching; dashed vertical lines mark the posterior predictive medians,
and numeric labels indicate posterior predictive quantiles. Text annotations give
the moment-matched half-normal scale for use as empirical prior guidance, rounded
to 1~decimal place for treatment effects and to 2~decimal places for interactions.
For odds ratios, risk ratios, hazard ratios, and rate ratios, $\tau^\ast$ is on
the log scale; for risk difference and standardised mean differences, $\tau^\ast$
is on the absolute scale}
\label{fig:heterogeneity-all-groups}
\end{figure}

Figure~\ref{fig:heterogeneity-all-groups} displays the predictive distributions of heterogeneity for treatment effects as well as treatment-by-subgroup interactions by outcome type.
For each outcome type, a histogram shows the predictive distribution as estimated using Markov chain Monte Carlo (MCMC), while the solid line illustrates a half-normal distribution whose scale is fitted to the data (by matching expectations, and rounding up to the next significant digit). 
Comparisons within an effect measure and its interaction—for example, between OR and ROR heterogeneity priors—share a common empirical basis because both are estimated from the same set of meta-analyses. Thus, any observed differences are not due to differences in the number of meta-analyses available for calibration. 
Each of the six figure panels draws on a different collection of meta-analyses with different clinical contexts, outcome distributions, and sizes. 
Figure~\ref{fig:heterogeneity-all-groups} shows that for each outcome type the predicted treatment-by-subgroup interaction heterogeneity is generally smaller than the corresponding treatment effect heterogeneity. Across effect measures, however, differences in the predictive distributions may reflect not only differences in scale and estimand, but also differences in the amount of data available for each type.

Table~\ref{tab:prior-comparison} shows the CDSR-based prior specifications derived here alongside the corresponding recommendations reported by 
R\"{o}ver \emph{et~al}.~[\cite{RoeverEtAl2021}], 
Rhodes \emph{et~al}.~[\cite{RhodesEtAl2015}], 
Turner \emph{et~al}.~[\cite{TurnerEtAl2015}]\, and 
Lilienthal \emph{et~al}.~[\cite{LilienthalEtAl2024}].
We focus on these comparators because they represent distinct and complementary approaches: the recommendations by Lilienthal \emph{et~al}.\ are likewise empirically derived from large-scale evidence syntheses, whereas the commonly used $\halfnormaldistn(0.5)$ priors discussed by R\"{o}ver et al.\ are motivated by general considerations of plausible between-study variability and are intended as broadly applicable weakly informative defaults rather than data-driven summaries. 

\begin{table}[!htbp]
\centering
\caption{Empirically informed prior recommendations for between-study heterogeneity, stratified by treatment effect scale, together with comparator prior proposals from the literature. The treatment-effect prior columns report prior distributions for the heterogeneity of overall treatment effects, including the new empirical Cochrane Database of Systematic Reviews (CDSR) priors derived in this work and previously proposed weakly informative priors (WIPs) or empirical priors. The interaction-prior columns report the corresponding new empirical CDSR priors for treatment-by-subgroup interaction heterogeneity. In the table, \(\mathrm{HN}\) denotes a half-normal distribution. Outcome scales include risk difference (RD), risk ratio (RR), hazard ratio (HR), odds ratio (OR), standardised mean difference (SMD), and incidence rate ratio (IRR). Interaction scales include difference of risk differences (DRD), ratio of risk ratios (RRR), ratio of hazard ratios (RHR), ratio of odds ratios (ROR), difference of standardised mean differences (DSMD), and ratio of incidence rate ratios (RIRR). The interaction priors are shown separately because they refer to heterogeneity in interaction effects and are therefore not directly comparable to priors for overall treatment-effect heterogeneity}
\label{tab:prior-comparison}

\scriptsize
\setlength{\tabcolsep}{3pt}
\renewcommand{\arraystretch}{1.15}

\begin{tabular}{@{} p{1.8cm} C{1.8cm} C{2cm} C{3cm} C{2.5cm} C{2cm}@{}}
\toprule
& \multicolumn{4}{c}{treatment effect prior}
& \multicolumn{1}{c}{interaction prior} \\
\cmidrule(lr){2-5}\cmidrule(l){6-6}

\shortstack{outcome /\\interaction\\scale}
& New CDSR \newline Figure \ref{fig:heterogeneity-all-groups}
& suggested WIPs\newline R\"{o}ver \emph{et~al}. [\cite{RoeverEtAl2021}]
& empirical CDSR\newline Rhodes / Turner [\cite{RhodesEtAl2015, TurnerEtAl2015}]
& empirical IQWiG\newline Lilienthal \emph{et~al}. [\cite{LilienthalEtAl2024}]
& New CDSR\newline Figure \ref{fig:heterogeneity-all-groups} \\
\midrule

RD / DRD
& \(\mathrm{HN}(0.1)\) 
&  
&  
&  
& \(\mathrm{HN}(0.05)\) \\

RR / RRR
& \(\mathrm{HN}(0.2)\) 
&  
&  
& \(\mathrm{HN}(0.1)\) 
& \(\mathrm{HN}(0.05)\) \\

HR / RHR
& \(\mathrm{HN}(0.3)\) 
&  
&  
& \(\mathrm{HN}(0.1)\) 
& \(\mathrm{HN}(0.03)\) \\

OR / ROR
& \(\mathrm{HN}(0.4)\) 
& \(\mathrm{HN}(0.5)\)\textsuperscript{a} 
& \(\lognormaldistn(-1.28,0.87^2)\)\textsuperscript{c}
& \(\mathrm{HN}(0.2)\) 
& \(\mathrm{HN}(0.07)\) \\

SMD / DSMD
& \(\mathrm{HN}(0.4)\) 
& \(\mathrm{HN}(0.5)\)\textsuperscript{a} 
& \(\logtdistn_{5}(-1.72,1.295)\)\textsuperscript{d}
& \(\mathrm{HN}(0.3)\) 
& \(\mathrm{HN}(0.14)\) \\

IRR / RIRR
& \(\mathrm{HN}(0.6)\) 
& \(\mathrm{HN}(0.5)\)\textsuperscript{b} 
&  
&  
& \(\mathrm{HN}(0.07)\) \\

\bottomrule
\end{tabular}

\vspace{0.75em}
\begin{minipage}{\linewidth}
\footnotesize
\smallskip
\noindent\textsuperscript{a} R\"{o}ver et al.\ use half-Normal(0.5) as a conservative
weakly informative choice in illustrative OR/SMD-type examples.\\
\textsuperscript{b} R\"{o}ver et al.\ use half-Normal(0.5) in a log-IRR example.\\
\textsuperscript{c} Turner et al.\ propose an empirical log-normal prior for
log-OR heterogeneity in a general healthcare setting.\\
\textsuperscript{d} Rhodes et al.\ propose an empirical log-\(t_5\) prior for
SMD heterogeneity.
\end{minipage}
\end{table}

Spiegelhalter \emph{et~al}.~[\cite{SpiegelhalterEtAl}] motivated heterogeneity categories for endpoints on a logarithmic scale by interpreting heterogeneity in terms of the difference between two randomly selected true study effects. On the log scale, this difference represents a contrast between study-specific treatment effects; exponentiated, it becomes a ratio. Larger heterogeneity thus implies larger multiplicative differences between study effects. The same reasoning applies to log-ROR interaction effects. In that setting, the difference between two randomly selected study-specific interaction effects corresponds, after exponentiation, to the ratio between two study-specific RORs. We use the same categories as a common reference scale in the
log-OR setting, comparing heterogeneity in OR treatment effects with
heterogeneity in ROR interaction effects on the corresponding log-ratio scale~[\cite{RoeverEtAl2021}]. Table~\ref{tab:spiegelhalter-prior-probabilities} reports the prior probability assigned by each log-OR heterogeneity prior (Empirical CDSR, Turner et al., R\"{o}ver et al., and IQWiG) to these heterogeneity categories. The empirical ROR interaction prior places most of its mass in the small-heterogeneity category, implying that large differences are less plausible.

\begin{table}[!htbp]
  \centering
  \caption{Prior probabilities assigned to heterogeneity intervals on the \(\tau\) scale. The categories follow the interpretation of log-odds-ratio heterogeneity given by Spiegelhalter et al. [\cite{SpiegelhalterEtAl}], including the ``small'' category given in Röver et al. [\cite{RoeverEtAl2021}]: small \((\tau < 0.1)\), reasonable \((0.1 \leq \tau < 0.5)\), fairly high \((0.5 \leq \tau < 1.0)\), and fairly extreme \((\tau \geq 1.0)\). The first four rows summarise treatment-effect heterogeneity priors for ORs from the literature, whereas the final row gives our proposed prior for interaction heterogeneity on the log-ROR scale, \(\mathrm{HN}(0.07)\). In the table, \(\mathrm{HN}\) denotes a half-normal distribution. The interval categories are used as a common log-ratio reference scale to facilitate comparison}
  \label{tab:spiegelhalter-prior-probabilities}

  \scriptsize
  \setlength{\tabcolsep}{2.5pt}
  \renewcommand{\arraystretch}{1.08}

  \begin{adjustbox}{max width=\textwidth}
  \begin{tabular}{@{} p{4.95cm} C{1.2cm} C{2cm} C{2cm} C{1.8cm} C{0.75cm} C{0.75cm} C{0.75cm} C{0.75cm} C{0.75cm}@{}}
    \toprule
    & \multicolumn{4}{c}{Probability}
    & \multicolumn{5}{c}{Summary} \\
    \cmidrule(lr){2-5}
    \cmidrule(lr){6-10}

    heterogeneity prior
    & \shortstack{small\\\((\tau < 0.1)\)}
    & \shortstack{reasonable\\\((0.1 \leq \tau < 0.5)\)}
    & \shortstack{fairly high\\\((0.5 \leq \tau < 1.0)\)}
    & \shortstack{fairly extreme\\\((\tau \geq 1.0)\)}
    & Mean
    & SD
    & 50\%
    & 95\%
    & 99\% \\
    \midrule

    \multicolumn{10}{@{}l}{\textit{treatment effect, log-OR scale}} \\

    Empirical CDSR, \(\mathrm{HN}(0.4)\)
    & 0.20 & 0.59 & 0.20 & 0.01
    & 0.32 & 0.24 & 0.27 & 0.78 & 1.03 \\

    Turner et al., \(\lognormaldistn(-1.28, 0.87^2)\)
    & 0.12 & 0.63 & 0.18 & 0.07
    & 0.41 & 0.43 & 0.28 & 1.16 & 2.10 \\

    R\"{o}ver et al., \(\mathrm{HN}(0.5)\)
    & 0.16 & 0.52 & 0.27 & 0.05
    & 0.40 & 0.30 & 0.34 & 0.98 & 1.29 \\

    Lilienthal et al., \(\mathrm{HN}(0.2)\)
    & 0.38 & 0.60 & 0.01 & 0.00
    & 0.16 & 0.12 & 0.13 & 0.39 & 0.52 \\[1ex]

    \multicolumn{10}{@{}l}{\textit{interaction, log-ROR scale}} \\

    Empirical CDSR, \(\mathrm{HN}(0.07)\)
    & 0.85 & 0.15 & 0.00 & 0.00
    & 0.06 & 0.04 & 0.05 & 0.14 & 0.18 \\

    \bottomrule
  \end{tabular}
  \end{adjustbox}
\end{table}

\subsection{Dependence on the number of meta-analyses}
Differences between the CDSR priors across effect measures arise mainly from the effect scale, since the same numerical value of heterogeneity has different implications on ratio, difference, and standardised scales. However, they also depend on the amount of information available  (number of meta-analyses $M$) for each effect measure (see Table \ref{tab:size-comparison}). Where many meta-analyses were included, such as for RR, OR, and SMD, the analysis provides stronger evidence against very large heterogeneity values and therefore better constrains the upper tail of the predictive distribution. 

To examine how the resulting predictive heterogeneity distributions depend on the amount of calibration data, we repeated the summarising-prior analysis using calibration sets of different sizes within each effect-measure stratum. This was intended to assess changes in distributional shape, which are driven by constraints on and uncertainty in the underlying scale parameter~$s$, and in particular, to investigate whether the differences observed in effect measures such as IRR and RD could be due to differing amounts of calibration data available. Specifically, for each effect measure, we estimated the posterior predictive heterogeneity distribution using subsets of size \(M = 5, 10, 20,\) and \(40\) meta-analyses, including the most recent, and compared these with the results obtained from the full set of eligible meta-analyses in that stratum. 

Table~\ref{tab:size-comparison} reports posterior predictive summaries across calibration subset sizes; the summaries are more variable for smaller calibration sets for treatment effect heterogeneity (\(\tau^\ast\)) and treatment-by-subgroup interaction heterogeneity (\(\tau^\ast_\gamma\)); this pattern is consistent with the arguments in Section~\ref{sec:Heterogeneityofinteractioneffects}. Within-trial interactions are more variable than treatment effect estimates. However, interaction heterogeneity tends to be smaller than overall treatment effect heterogeneity, while at the same time being harder to estimate because interaction estimates typically have larger standard errors. Consequently, when the number of calibration meta-analyses~\(M\) is small, the posterior predictive distribution for interaction heterogeneity may exhibit larger medians or upper quantiles, reflecting weak empirical constraint rather than genuinely greater heterogeneity. As \(M\) increases, this uncertainty is reduced, and the smaller magnitude of interaction heterogeneity becomes more clearly apparent. This pattern is also reflected in the mean- and median-ratio columns in Table~\ref{tab:size-comparison}, which indicate that interaction heterogeneity is typically smaller on average even when its posterior predictive distribution remains comparatively diffuse for small~\(M\).

\begin{table}[!htbp]
  \centering
  \footnotesize
  \caption{Posterior predictive heterogeneity summaries comparing treatment effect~($\tau^\ast$) and treatment-by-subgroup interaction heterogeneity~($\tau_\gamma^\ast$) for each effect measure. Rows are grouped by effect measure and ordered by calibration subset size ($M=5$, $M=10$, $M=20$, $M=40$, all analyses). For both heterogeneity parameters ($\tau^\ast$, $\tau_\gamma^\ast$), columns report mean, Std.dev., 50\%, 95\%, and 99\% quantiles. The last two columns report interaction-to-treatment effect ratios (median and mean ratio), where values less than~1 indicate smaller interaction heterogeneity on that summary}
  \label{tab:size-comparison}
  \resizebox{\textwidth}{!}{%
  \begin{tabular}{@{}llrrrrrrrrrrrr@{}}
    \toprule
    & & \multicolumn{5}{c}{Overall effect ($\tau$)} & \multicolumn{5}{c}{Interaction (\(\tau_\gamma\))} & \multicolumn{2}{c}{Comparison ($\tau_\gamma / \tau$)} \\
    \cmidrule(lr){3-7}\cmidrule(lr){8-12}\cmidrule(lr){13-14}
    Outcome & Data & Mean & Std.dev. & 50\% & 95\% & 99\% & Mean & Std.dev. & 50\% & 95\% & 99\% & Median ratio & Mean ratio \\
    \midrule
    \multirow{5}{*}{RD} & $M = 5$ & 0.04 & 0.05 & 0.02 & 0.13 & 0.26 & 0.10 & 0.17 & 0.05 & 0.36 & 0.73 & 3.06 & 2.83 \\
     & $M = 10$ & 0.03 & 0.03 & 0.02 & 0.08 & 0.13 & 0.03 & 0.04 & 0.02 & 0.11 & 0.19 & 1.00 & 1.19 \\
     & $M = 20$ & 0.04 & 0.03 & 0.03 & 0.10 & 0.15 & 0.01 & 0.01 & 0.00 & 0.03 & 0.06 & 0.14 & 0.22 \\
     & $M = 40$ & 0.03 & 0.02 & 0.02 & 0.07 & 0.11 & 0.00 & 0.00 & 0.00 & 0.01 & 0.03 & 0.09 & 0.14 \\
     & All analyses ($M = 44$) & 0.06 & 0.04 & 0.05 & 0.14 & 0.18 & 0.04 & 0.03 & 0.04 & 0.11 & 0.15 & 0.77 & 0.77 \\
    \midrule
    \multirow{5}{*}{RR} & $M = 5$ & 0.30 & 0.45 & 0.18 & 0.95 & 2.08 & 0.63 & 0.89 & 0.37 & 2.10 & 4.18 & 2.10 & 2.10 \\
     & $M = 10$ & 0.13 & 0.14 & 0.09 & 0.41 & 0.64 & 0.14 & 0.19 & 0.08 & 0.53 & 0.89 & 0.84 & 1.09 \\
     & $M = 20$ & 0.16 & 0.13 & 0.13 & 0.42 & 0.60 & 0.07 & 0.09 & 0.04 & 0.24 & 0.42 & 0.29 & 0.43 \\
     & $M = 40$ & 0.13 & 0.10 & 0.10 & 0.33 & 0.46 & 0.07 & 0.09 & 0.04 & 0.25 & 0.43 & 0.42 & 0.60 \\
     & All analyses ($M = 2{,}025$) & 0.20 & 0.15 & 0.17 & 0.49 & 0.65 & 0.04 & 0.05 & 0.02 & 0.13 & 0.21 & 0.13 & 0.19 \\
    \midrule
    \multirow{5}{*}{OR} & $M = 5$ & 0.77 & 1.21 & 0.39 & 2.82 & 5.97 & 0.90 & 1.44 & 0.38 & 3.52 & 7.06 & 0.98 & 1.16 \\
     & $M = 10$ & 0.38 & 0.42 & 0.24 & 1.18 & 1.98 & 0.41 & 0.58 & 0.20 & 1.51 & 2.70 & 0.83 & 1.09 \\
     & $M = 20$ & 0.32 & 0.32 & 0.23 & 0.97 & 1.45 & 0.20 & 0.27 & 0.11 & 0.74 & 1.22 & 0.49 & 0.63 \\
     & $M = 40$ & 0.37 & 0.30 & 0.30 & 0.96 & 1.30 & 0.09 & 0.12 & 0.05 & 0.32 & 0.55 & 0.16 & 0.24 \\
     & All analyses ($M = 346$) & 0.33 & 0.26 & 0.28 & 0.82 & 1.13 & 0.06 & 0.06 & 0.04 & 0.19 & 0.31 & 0.13 & 0.18 \\
    \midrule
    \multirow{5}{*}{IRR} & $M = 5$ & 0.24 & 0.45 & 0.12 & 0.80 & 1.87 & 0.24 & 0.43 & 0.11 & 0.91 & 1.85 & 0.94 & 1.02 \\
     & $M = 10$ & 0.17 & 0.19 & 0.11 & 0.54 & 0.88 & 0.13 & 0.17 & 0.07 & 0.45 & 0.81 & 0.60 & 0.72 \\
     & $M = 20$ & 0.26 & 0.23 & 0.20 & 0.72 & 1.05 & 0.10 & 0.13 & 0.05 & 0.35 & 0.59 & 0.26 & 0.38 \\
     & $M = 40$ & --- & --- & --- & --- & --- & --- & --- & --- & --- & --- & --- & --- \\
     & All analyses ($M = 28$) & 0.44 & 0.36 & 0.36 & 1.14 & 1.62 & 0.06 & 0.08 & 0.03 & 0.21 & 0.38 & 0.09 & 0.13 \\
    \midrule
    \multirow{5}{*}{HR} & $M = 5$ & 0.08 & 0.13 & 0.04 & 0.30 & 0.53 & 0.24 & 0.36 & 0.12 & 0.85 & 1.86 & 2.90 & 2.89 \\
     & $M = 10$ & 0.06 & 0.08 & 0.04 & 0.22 & 0.36 & 0.12 & 0.17 & 0.06 & 0.45 & 0.84 & 1.83 & 1.98 \\
     & $M = 20$ & 0.05 & 0.06 & 0.03 & 0.18 & 0.32 & 0.08 & 0.11 & 0.04 & 0.30 & 0.54 & 1.50 & 1.62 \\
     & $M = 40$ & 0.15 & 0.16 & 0.09 & 0.48 & 0.72 & 0.48 & 0.45 & 0.36 & 1.38 & 1.97 & 3.97 & 3.24 \\
     & All analyses ($M = 144$) & 0.25 & 0.19 & 0.21 & 0.61 & 0.81 & 0.03 & 0.03 & 0.02 & 0.09 & 0.15 & 0.08 & 0.11 \\
    \midrule
    \multirow{5}{*}{SMD} & $M = 5$ & 0.46 & 0.62 & 0.28 & 1.54 & 2.72 & 0.41 & 0.60 & 0.20 & 1.51 & 2.87 & 0.72 & 0.88 \\
     & $M = 10$ & 0.23 & 0.23 & 0.17 & 0.67 & 1.07 & 0.18 & 0.23 & 0.10 & 0.61 & 1.12 & 0.62 & 0.77 \\
     & $M = 20$ & 0.64 & 0.53 & 0.52 & 1.69 & 2.37 & 0.04 & 0.06 & 0.02 & 0.16 & 0.29 & 0.04 & 0.07 \\
     & $M = 40$ & 0.42 & 0.34 & 0.35 & 1.06 & 1.45 & 0.03 & 0.04 & 0.02 & 0.12 & 0.22 & 0.05 & 0.08 \\
     & All analyses ($M = 475$) & 0.30 & 0.23 & 0.26 & 0.75 & 0.99 & 0.11 & 0.08 & 0.09 & 0.27 & 0.36 & 0.35 & 0.36 \\
     \bottomrule
  \end{tabular}%
  }
\end{table}

\section{Example application: Intravenous iron in heart failure hospitalisations}
\label{sec:MotivatingExample}

Anker et al.\ (2025) report a systematic review and meta-analysis summarising the efficacy and safety of intravenous iron from six randomised trials including a total of 7{,}175~patients~[\cite{Anker2025}]; the data are readily available in [\cite{metadat-package}].
The primary endpoint is a composite of total (first and recurrent) heart failure hospitalisations and cardiovascular mortality, 
yielding incidence rate ratios (IRRs) as treatment effects. The interaction of interest compares the effect of intravenous iron between women and men as an interaction on the log-IRR scale, corresponding to a ratio of incidence rate ratios (RIRR) on the original scale (Section~\ref{sec:Heterogeneityofinteractioneffects}). A key aspect of this example is the choice of prior for the interaction heterogeneity parameter~\(\tau_\gamma\), which may substantially affect inference given the small number of available studies. In the original analysis, a half-normal prior with scale \(0.5\) (\(\halfnormaldistn(0.5)\)) was used, reflecting a commonly adopted weakly informative specification for treatment effect heterogeneity (see Table \ref{tab:prior-comparison}). Sensitivity analyses additionally considered a broader prior (\(\halfnormaldistn(1.0)\)), and a narrower prior (\(\halfnormaldistn(0.1)\)), representing more conservative and more optimistic assumptions about between-study variability, respectively. 

\begin{figure}
 \centering
 \includegraphics[width=\linewidth]{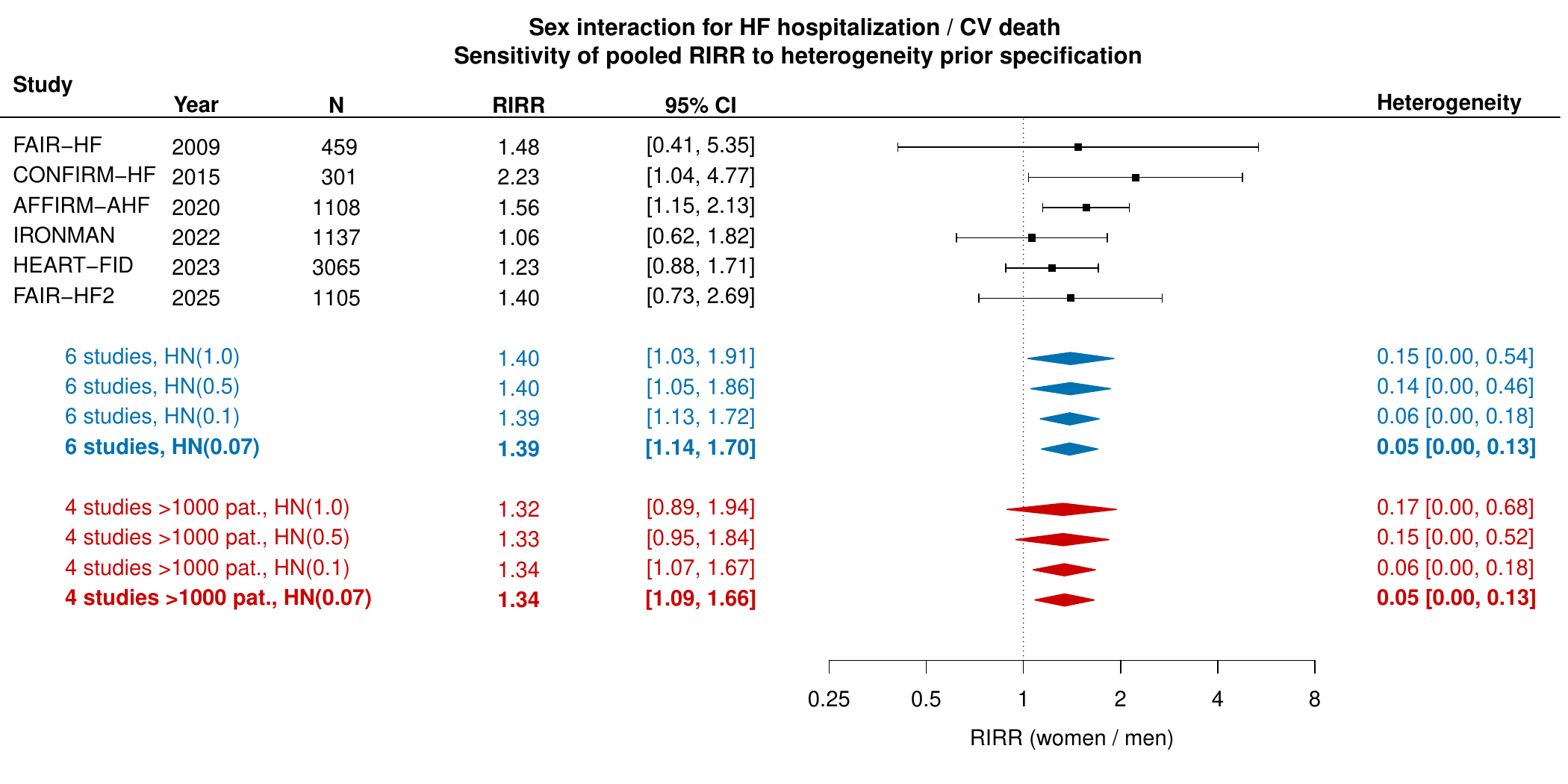}
\caption{Pooled treatment-by-sex interaction for heart failure (HF) hospitalization or cardiovascular (CV) death, expressed as the ratio of incidence rate ratios (RIRR, women/men), under alternative prior choices for the interaction heterogeneity $\tau_\gamma$. Results are shown for all six studies and for a sensitivity analysis restricted to the four most recent and largest studies, each including more than 1,000~patients. Pooled posterior medians with 95\% credible intervals are reported under half-Normal priors with scale parameters 1.0, 0.5, 0.1, and 0.07. Bold font face indicates the new empirical prior proposal in Figure \ref{fig:heterogeneity-all-groups}}
 \label{fig:sex-subgroup}
\end{figure}

Figure~\ref{fig:sex-subgroup} illustrates the study-specific and pooled estimates of the treatment-by-sex interaction, expressed as the RIRR for women versus men, under different heterogeneity prior specifications: the three priors used in the original analysis, and a $\halfnormaldistn(0.07)$~prior that would be suggested based on the analyses in Section~\ref{sec:Empiricaldata}. Besides the primary analysis including all six studies, a sensitivity analysis considered only the four most recent and largest studies (including at least 1,000~patients: AFFIRM-AHF, IRONMAN, HEART-FID, FAIR-HF2).

In the present example including only a few studies, interaction heterogeneity is only weakly identified by the available data, and consequently prior specifications are influential for the posterior. When all six studies are included, the choice of prior has little effect on the RIRR estimate, only the credible interval (CrI) width changes, while in all cases the RIRR of~$1.0$ (indicating no interaction) remains outside the CrI.
In the sensitivity analysis with only four studies considered, the situation changes; the estimated RIRR is slightly lower, but depending on the specified prior, the CrI only excludes an RIRR of~$1.0$ for the priors with scales~$\leq 0.1$. Utilising external evidence on likely interaction heterogeneity magnitudes in this case makes it possible to arrive at the same qualitative conclusion based on a smaller data set.

\section{Discussion}
\label{sec:Discussion}

We have derived empirical prior distributions for treatment-by-subgroup interaction heterogeneity from more than 3,000 CDSR meta-analyses, extending the empirical prior literature to the interaction scale. Across all effect measures considered, interaction heterogeneity was found to be substantially smaller than treatment effect heterogeneity.

Heterogeneity is defined in meta-analysis of interventions as variability in intervention effects being evaluated across studies, attributable to clinical or methodological diversity and reflected in observed effects that differ by more than would be expected by chance alone~[\cite{CochraneHandbookV6}]. In the interaction setting, however, heterogeneity pertains to variation across studies in how intervention effects differ between subgroups, rather than to variation in the intervention effects themselves. This distinction provides a plausible explanation for the smaller empirical interaction heterogeneity observed here: study-level factors that shift treatment effects similarly in both subgroups contribute to treatment-effect heterogeneity but may cancel out in within-trial interaction contrasts. Interaction heterogeneity would therefore arise only from factors that modify the subgroup \emph{contrast} itself, despite the stratification effectively implemented by comparing subgroups within often randomised studies. The presence of such ``interaction modifier'' effects of substantial magnitude may therefore be less readily expected, although this will depend on the clinical and methodological context. This interpretation should be distinguished from the separate issue that interaction estimates are less precise and heterogeneity on the interaction scale is harder to estimate. Interaction heterogeneity is generally estimated with substantial uncertainty, particularly in sparse settings; this is reflected in the interaction-scale summaries in Table~\ref{tab:size-comparison} and explained by the precision argument in Section~\ref{sec:Heterogeneityofinteractioneffects}.

The results are consistent with previously published work by Rhodes \emph{et~al}. [\cite{RhodesEtAl2015}],  Turner \emph{et~al}. [\cite{TurnerEtAl2015}], R\"{o}ver \emph{et~al}. [\cite{RoeverEtAl2021}], and Lilienthal \emph{et~al}. [\cite{LilienthalEtAl2024}], and extend that literature to the interaction scale.
When considering such empirically motivated prior distributions, keeping in mind the original application context is crucial.
Earlier investigations of between-study heterogeneity [\cite{RhodesEtAl2015,TurnerEtAl2015}] had been based on the CDSR, where Cochrane reviews are known to often employ relatively wide inclusion criteria.
Lilienthal \emph{et~al}. [\cite{LilienthalEtAl2024}] on the other hand considered a regulatory context and derived empirical priors from IQWiG data, explicitly noting that their setting is more restrictive; consequently, they found less heterogeneity in their data. The results have practical implications for Bayesian random-effects meta-analysis of interactions. In sparse settings, interaction heterogeneity is weakly identified and prior assumptions can materially influence inference~[\cite{FriedeRoeverWandelNeuenschwander2017a, BenderEtAl2018}], as was also illustrated by the motivating example in Section~\ref{sec:MotivatingExample}.
Empirically calibrated priors can complement theoretical arguments for motivating (weakly) informative specifications, or may serve as an external validation.

Some limitations should be noted.
First, this calibration is restricted to pairwise subgroup comparisons; extension to settings with multiple treatments or multi-level subgroup structures~[\cite{PhillippoEtAl2020}] would require consideration of more complex models.
Second, interaction estimates are derived from published subgroup summaries and may therefore be affected by selective reporting, by variation in subgroup definitions across trials, and by multiplicity in the calibration database~[\cite{LilienthalEtAl2024}]. In particular, the same underlying trials, endpoints, or closely related analyses may contribute more than once, so the reference set is not necessarily fully independent (although we would not expect substantial biases from such model violations).

\appendix

\clearpage
\begin{appendix}
\end{appendix}
\clearpage

\begin{Backmatter}

\paragraph{Funding Statement}
  Support from the \emph{Volkswagen Stiftung} (project ``Bayesian and nonparametric statistics -- Teaming up two opposing theories for the benefit of prognostic studies in \mbox{COVID-19}''), the German Centre for Cardiovascular Research \emph{(Deutsches Zentrum für Herz-Kreislauf-Forschung e.V., DZHK)}, and the \emph{Deutsche Forschungsgemeinschaft (DFG)} is gratefully acknowledged (DZHK grant number \mbox{81Z0300108}; DFG grant number \mbox{FR~3070/3-2}, project number 413270747).

\paragraph{Competing Interests}
  The authors have declared no conflicts of interest.

\paragraph{Data Availability Statement}
  Data from the example application (Section~\ref{sec:MotivatingExample}) are available in the \texttt{metadat} R~package [\cite{metadat-package}].


\paragraph{Author Contributions}
Conceptualization: R.P; C.R.;T.F. Methodology: R.P; C.R.;T.F. 
Writing original draft: R.P. All authors approved the final submitted draft.



\printbibliography

\end{Backmatter}
\end{document}